\documentclass{ifacconf}

\usepackage{graphicx}      
\usepackage{amsmath, amssymb, amsfonts}
\usepackage{natbib}        
\usepackage{todonotes}
\usepackage{tikz}
\usepackage{pgfplots}
\pgfplotsset{compat=newest}
\usepgfplotslibrary{groupplots} 
\usepgfplotslibrary{statistics}
\usetikzlibrary{patterns}

\pgfplotsset{
  lakeActual/.style={blue, thick},
  lakeTarget/.style={green, dashed, very thick},
  lakeMax/.style={red, dotted, thick},
  lakeMin/.style={cyan, dotted, thick},
  lakeInflow/.style={const plot, blue, dashed, very thick},
  lakeOutflow/.style={const plot, red, thick},
  lakeDepth/.style={ybar, bar width=2pt, fill, draw=none},
  lakeMean/.style={red, dashed, very thick},
}

\begin{document}
\begin{frontmatter}

\title{A characteristic function framework for chance constraint programming in stochastic model predictive control\thanksref{footnoteinfo}} 

\thanks[footnoteinfo]{This research was supported by the ELLIIT strategic research area.}

\author[First]{Yuwei Ying} 
\author[First]{Johan Löfberg} 
\author[First]{Anders Hansson}

\address[First]{Dept. of Electrical Engineering, Linköping University, SE+581 83 Linköping, Sweden (e-mail: yuwei.ying@liu.se)}

\begin{abstract}                
The computation of chance constraints in stochastic model predictive control is often numerically challenging due to the non-Gaussian nature of the disturbances. To overcome this problem, we propose an optimization computational framework applicable to non-Gaussian disturbances. This framework employs a numerical inversion method, utilizing the characteristic function of the disturbance distribution to compute the probability in the chance constraint as well as its gradient. To improve efficiency, it vectorizes integral points and reuses intermediate computations in Gauss-Kronrod quadrature. The framework is implemented within the YALMIP toolbox to perform chance constraint calculations for arbitrary non-Gaussian disturbances, applicable to both single-component distributions and mixture models. It allows the user to simply specify a distribution type and its parameters for the disturbance and directly compute the probability and its gradient to solve the optimization problem. The method is validated through a numerical example of a stochastic model predictive control application.
\end{abstract}


\end{frontmatter}

\section{Introduction}
Model predictive control (MPC) is a model-based optimal control technique that is applied in many fields such as industrial process control \citep{Qin:03} and energy systems \citep{Ras:14}. It has three core steps: prediction, optimization and receding horizon implementation. It uses a mathematical model of the system to predict future behavior and obtains the optimal control sequence over a finite prediction horizon through optimization algorithms. Finally, only the first control action is executed, and the process repeats in a receding horizon manner. \par
As a deterministic optimization method, classical MPC assumes a known system model and cannot directly handle model uncertainty or external random disturbances. To address this problem, robust MPC (RMPC) emerged as a natural extension of classical MPC. It introduces a min-max optimization framework \citep{Bem:99} that considers the worst-case scenario of bounded uncertainty. In this framework, the controller optimizes its performance by assuming the most unfavorable disturbance sequence within a predefined set of uncertainties, thereby ensuring constraint satisfaction under all admissible disturbances. However, this also leads to over-conservatism \citep{Joe:20} since it optimizes for the worst-case scenario at each step, sacrificing performance even when actual disturbances are minor or infrequent. To some extent, this can be mitigated by disturbance feedback \citep{Lof:03}. Furthermore, the computational complexity of solving min-max optimization problems increases rapidly with system dimension \citep{De:06}, making real-time implementation difficult to achieve. \par
To cope with these limitations, stochastic MPC (SMPC) was subsequently developed. SMPC models disturbances as random variables with known or estimated probability distributions and reformulates hard constraints as chance constraints \citep{Old:08}, permitting violations of these constraints with an acceptable probability. This probabilistic relaxation significantly reduces conservatism while maintaining statistical safety guarantees. \par
In early studies, SMPC typically employed the Gaussian assumption \citep{Bla:09} for analytical tractability. However, uncertainty in the real world often exhibits non-Gaussian characteristics. To eliminate reliance on the Gaussian distribution, a safe approximation \citep{Tan:24} based on first and second moment information has been proposed. The chance constraint can be reformulated as a second-order cone constraint, making it computationally efficient. However, it is highly conservative and performs poorly for skewed or heavy-tailed disturbances. In addition, the scenario-based approach \citep{Geo:14} is also a widely used method. This method uses sampled disturbance scenarios to construct scenario constraints and statistical theory to guarantee reliability and feasibility. However, it requires a large number of samples to provide strict probabilistic guarantees. The number of scenarios increases rapidly with the problem dimension and risk requirements, resulting in large computational load and poor real-time performance. The conditional Value at Risk (CVaR) method \citep{Kel:21} replaces the chance constraint with the CVaR constraint, which captures the mean loss over the worst $\varepsilon$ tail of the distribution. It is a convex constraint and can be estimated by sampling, but may not be equivalent to the original chance constraint.\par
Although existing approaches provide useful solutions for handling chance constraints in SMPC, their applicability is often limited by specific assumptions, such as Gaussian disturbances and sampling-based approximations. Motivated by this, we develop a general framework that offers a tractable way to handle chance constraints for a wider range of non-Gaussian disturbances. Furthermore, we reduce computational cost through vectorized calculations and data reuse strategies during integration.
\subsubsection{Notation}The set of nonnegative integers and the set of real numbers are denoted by $\mathbb{Z}_+$ and $\mathbb{R}$, respectively. The $n$-dimensional Euclidean space is denoted by $\mathbb{R}^n$. We use $P(\cdot)$, $F(\cdot)$, $p(\cdot)$ and $\phi(\cdot)$ to denote the probability of an event, the cumulative distribution function (CDF), the probability density function (PDF) and the characteristic function (CF), respectively. The transpose of a vector or a matrix is denoted by $(\cdot)^\top$. The operators $\text{Im}[\cdot]$ and $\text{Re}[\cdot]$ denote the imaginary and real parts, respectively. $I_n$ denotes the $n\times n$ identity matrix. The operator $\mathbb{E}[\cdot]$ denotes the statistical expectation.

\section{Preliminaries and Problem Statement}
This section first introduces the structure of stochastic MPC, then presents the standard form of general chance constraints. Finally, we discuss the computational challenges associated with handling chance constraints, demonstrating the necessity of a unified framework.

\subsection{Stochastic MPC background}
In this paper, we consider a discrete-time linear time-invariant system with random disturbance, given by
\begin{equation}
    x_{k+1} = Ax_k + Bu_k + Gw_k, \label{eqDynamics}
\end{equation}
where $k\in\mathbb{Z}_+$, $x_k\in\mathbb{R}^{n_x}$ is the system state, $u_k\in\mathbb{R}^{n_u}$ is the control input, $w_k\in\mathbb{R}^{n_w}$ is the additive stochastic disturbance, and $A,B,G$ are all known matrices. \par
The control objective of SMPC involves minimizing some performance measure over a prediction horizon, subject to the system dynamics as well as the input and state constraints. However, it is generally not possible to satisfy hard constraints on the input or the state due to the uncertainty in the random $w_k$. Therefore, chance constraints are introduced to replace hard constraints with probabilistic constraints.

\subsection{Chance constraint formulation}
To analyze the computation of chance constraints independently of the SMPC setting, we consider a generic affine (in $w$) chance constraint of the form
\begin{equation}
    P(q(z)+g^\top(z)w\leq0)\geq\gamma, \label{eqChanceCons}
\end{equation}
where $z\in\mathbb{R}^n$ are the decision variables, $w\in\mathbb{R}^m$ are random variables with mutually independent elements. The functions $q(\cdot):\mathbb{R}^n\rightarrow\mathbb{R}$ and $g(\cdot):\mathbb{R}^n\rightarrow\mathbb{R}^m$ are known functions, with $g(z)=[g_1(z),\cdots,g_m(z)]^\top$, and $\gamma\in[0,1]$ denotes the prescribed probability level, possibly also a decision variable. \par
This kind of representation isolates the essential structure of the chance constraint without relying on the notation of SMPC. In light of \eqref{eqChanceCons}, we next discuss the difficulties involved in computing the chance constraint. The connection to the SMPC formulation will be re-established later in the numerical example.

\subsection{Challenge of chance constraint computation}
To compute the probability $P(\cdot)$ in \eqref{eqChanceCons}, a closed-form CDF of $w$ is often used when available. However, such a form is generally limited to simple distributions. Since CFs are available in closed form for a wider range of distributions, we instead use numerical inversion based on the CF. For instance, when $w_j$ is normally distributed, we can define an auxiliary variable as follows
\begin{equation}
    \lambda=g^\top(z)w=\sum_{j=1}^{m}g_j(z)w_j, \label{eqZ}
\end{equation}
where $\lambda$ is a linear combination of $w_j$, which means $\lambda$ also follows a Gaussian distribution. In this case, the probability in \eqref{eqChanceCons} can be expressed as the CDF of $\lambda$
\begin{equation}
    F_{\lambda}\big(-q(z)\big)=P\big(\lambda\leq -q(z)\big), \label{eqCDF}
\end{equation}
which clearly depends on the decision variable $z$ through $-q(z)$.
Under this assumption, computing the CDF of $\lambda$ becomes straightforward. \par
However, this cannot be extended to all the distributions since only stable families \citep{Bor:05} such as Gaussian and Cauchy distributions are invariant under linear transformations. For cases where $w_j$ follows other distributions, a closed-form distribution of $\lambda$ is rarely available, and conventional computation of CDF becomes intractable. \par
Furthermore, in some practical applications, each independent component $w_j$ may arise from different modes and is therefore modeled by a mixture rather than a single-component distribution. Assuming $w_j$ consists of $R_j$ components, the PDF of $w_j$ takes the form
\begin{equation}
    p(w_j) = \sum_{r=1}^{R_j}\mu_{j,r}p_{j,r}(w_j),
\end{equation} 
where $\mu_{j,r}$ is the weight for the $r$-th mixture component in $w_j$. By independence, the joint PDF is
\begin{equation}
    p(w) = \prod_{j=1}^{m}p(w_j)=\prod_{j=1}^{m}\left(\sum_{r=1}^{R_j}\mu_{j,r}p_{j,r}(w_j)\right), \label{eqPw}
\end{equation}
This implies that the joint distribution is a mixture with $\prod_{j=1}^{m}{R_j}$ components, which is obtained by taking all combinations of the marginal components. \par
According to the distribution transformation formula \citep{Cas:02}, the PDF of $\lambda$ can be written as
\begin{equation}
    p(\lambda)=\int\delta(\lambda-g^\top(z)w)p(w)dw,
\end{equation}
where $\delta(\cdot)$ denotes the Dirac delta function. When $p(w)$ is expressed as a product of mixture distributions over each component, it follows that $p(\lambda)$ remains a mixture distribution, consisting of $\prod_{j=1}^{m}{R_j}$ components. In this case, a large value of either $R_j$ or $m$ will lead to an exponential increase of the computation burden.  \par
To address these two limitations mentioned above, we propose a computationally efficient framework applicable to both the single-component distribution and the mixture model for computing the probability in \eqref{eqCDF}. The details of the approach are presented in the following section.

\subsection{The bigger picture}

Extending the currently available modelling framework in YALMIP \citep{Lof:04} for deterministic uncertain models \citep{Lof:12} outside the standard linear Gaussian case has previously been severely hindered by the lack of computationally tractable methods, although some strategies have been implemented using exponential cone approximations \citep{Bar:25}. The computational framework outlined in this paper is thus part of an ongoing effort to add general support for stochastic programming in YALMIP \citep{Lof:04} for much larger classes of models.

\section{Computational framework for chance-constrained optimization}
In this section, we will first present the derivation of the probability expression obtained via numerical inversion, followed by a description of principal strategies used in our framework.

\subsection{Probability evaluation and its gradient}
As indicated by equation \eqref{eqCDF}, the task of computing the probability can be reformulated as the evaluation of the CDF $F_{\lambda}\big(-q(z)\big)$. We also require its gradient with respect to $z$, which provides sensitivity information for gradient-based optimization. For the sake of convenience, we denote $F_{\lambda}\big(-q(z)\big)$ by $\beta(z)$. \par
As noted earlier, there are cases where the linear combination makes it difficult to obtain a closed-form expression for the PDF of $\lambda$. In other cases, even when $p(\lambda)$ is available analytically, its complexity hinders the derivation of the antiderivative. However, for a random variable $X$, the CDF $F_X(x)$ and the PDF $p_X(x)$ are given by
\begin{equation}
    F_X(x)=P(X\leq x)=\frac{1}{2}-\frac{1}{\pi}\int_0^\infty \frac{\text{Im}[e^{-itx}\phi_X(t)]}{t}dt, \label{eqGILCDF}
\end{equation}
\begin{equation}
    p_X(x)=\frac{1}{\pi}\int_0^\infty \text{Re}[e^{-itx}\phi_X(t)]dt, \label{eqGILPDF}
\end{equation}
where $\phi_X(t)$ is the CF of $X$. Once the CF of $X$ is available, its CDF and PDF can be obtained via \eqref{eqGILCDF} and \eqref{eqGILPDF}, with \eqref{eqGILCDF} known as the \cite{Gil:51} inversion formula.\par
For independent random variables, the CF of their sum is given by the product of their individual CFs. When the distribution of $w_j$ is specified, the CF of $\lambda$ in \eqref{eqZ} can be expressed as
\begin{equation}
    \phi_{\lambda}(t;z) = \prod_{j=1}^{m}\phi_{w_j}(g_j(z)t). \label{eqjointCF}
\end{equation}
 With \eqref{eqCDF} and \eqref{eqGILCDF}, the CDF $\beta(z)$ can be expressed as 
\begin{equation}
    \beta(z)=F_\lambda\big(-q(z)\big)=\frac{1}{2}-\frac{1}{\pi}\int_0^\infty \frac{\text{Im}[e^{itq(z)}\phi_\lambda(t;z)]}{t}dt. \label{eqBeta}
\end{equation}
Similarly, the gradient w.r.t. $z$ can be expressed via \eqref{eqGILPDF} as 
\begin{equation}
    \begin{split}
        & \nabla_z{\beta(z)} = -\nabla_z{q(z)}p_\lambda(-q(z)) \\
        & -\frac{1}{\pi}\int_0^\infty \frac{\text{Im}[e^{itq(z)}\nabla_z\phi_\lambda(t;z)]}{t}dt, \label{eqDerivative}
    \end{split}
\end{equation}
where $p_\lambda(-q(z))=\frac{1}{\pi}\int_0^\infty \text{Re}[e^{itq(z)}\phi_\lambda(t;z)]dt$ and $p_\lambda(\cdot)$ is the PDF of $\lambda$.
Assuming that $\phi_{w_j}(g_j(z)t)\neq 0$ at the quadrature nodes, the gradient of $\phi_\lambda(t;z)$ can be expanded as
\begin{equation}
    \begin{split}
        & \nabla_z\phi_\lambda(t;z) = \nabla_z\left( \prod_{j=1}^{m}\phi_{w_j}(g_j(z)t)\right) \\
        & = \phi_\lambda(t;z)\sum_{j=1}^{m}\frac{t\phi_{w_j}'(g_j(z)t)}{\phi_{w_j}(g_j(z)t)}\nabla_z{g_j(z)}. \label{eqPartialD}
    \end{split}
\end{equation}
From \eqref{eqBeta} and \eqref{eqDerivative}, we obtain the CDF and its gradient, where $w_j$ is subject to a certain single-component distribution. We next extend them to mixture models.\par
For a random variable $w_j$ modeled by a mixture distribution, the CF satisfies the following relation
\begin{equation}
    \phi_{w_j}(g_j(z)t)=\sum_{r=1}^{R_j}\mu_{j,r}\phi_{j,r}(g_j(z)t).
\end{equation}
Thus, the equation in \eqref{eqPartialD} can be written as
\begin{equation}
    \begin{split}
        & \nabla_z\phi_\lambda(t;z) = \\
        & \phi_\lambda(t;z)\sum_{j=1}^{m}\frac{\sum_{r=1}^{R_j}\mu_{j,r}t\phi_{j,r}'(g_j(z)t)}{\phi_{w_j}(g_j(z)t)}\nabla_z{g_j(z)}. \label{eqPDforMixture}
    \end{split}
    \end{equation}
By substituting the above two equations into \eqref{eqBeta}-\eqref{eqDerivative}, one can obtain $\beta(z)$ and $\nabla_z{\beta(z)}$ for the mixture model case. It can be seen in \eqref{eqPDforMixture}, substituting the CF reduces the evaluation cost to $\sum_{j=1}^{m}R_j$, which is a substantial reduction compared to $\prod_{j=1}^{m}R_j$ in \eqref{eqPw}. From this perspective, the Gil-Pelaez inversion formula facilitates the computationally efficient evaluation of chance constraints under general disturbance distributions.

\subsection{Optimized integration scheme}
To ensure the required level of computational accuracy, the Gauss-Kronrod 7-15 quadrature is utilized to build an adaptive numerical integration procedure to compute $\beta(z)$ and $\nabla_z{\beta(z)}$. However, this requires multiple evaluations of the integrand, as it uses 15 nodes in each subdivision. Therefore, another key point of this paper is to lower the computational burden when dealing with numerical integration. \par
To obtain a unified expression for both single-component and mixture cases, define
\begin{equation}
    \xi(z,t)=\sum_{j=1}^m\alpha_j(z,t)\nabla_z g_j(z)
\end{equation}
where $\alpha_j(z,t)$ denotes the corresponding scalar factor in \eqref{eqPartialD} or \eqref{eqPDforMixture}. Then $\nabla_z\phi_\lambda(t;z) =\phi_\lambda(t;z)\xi(z,t)$, and the gradient of $\beta(z)$ can be written as
\begin{equation}
    \begin{split}
        & \nabla_z{\beta(z)} = -\nabla_z{q(z)}p_\lambda(-q(z)) - \\
        & \frac{1}{\pi}\int_0^\infty \frac{\text{Im}[e^{itq(z)}\phi_\lambda(t;z)\xi(z,t)]}{t}dt. \label{eqGeneral}
    \end{split}
\end{equation}
Now the task is to calculate \eqref{eqBeta} and \eqref{eqGeneral}. Here, several strategies are employed to improve efficiency.

\subsubsection{1. Efficient exploitation of vectorization} In each integration cycle, a strategy of structured data processing and vectorized computation is employed to reduce the computational cost. To be more precise, all 15 integration nodes $t$ are constructed as a vector. Instead of evaluating each node individually, the entire vector is passed to the predefined characteristic function and its gradient function module. This ensures that the function values of these 15 nodes are generated in parallel within a single operation, reducing the overhead of function calls and the latency of loop iterations.
\subsubsection{2. Result matrixization and reuse across quadrature rules}
After the function calculation, the output is systematically stored in an intermediate data structure in matrix form. This storage mode not only facilitates data management and access, but also allows subsequent calculation procedure to access the corresponding values directly. Since Gauss-7 and Kronrod-15 quadrature rules rely on the same characteristic function and gradient evaluations during the calculation, differing only in the specific integration weighting schemes, only a single function evaluation is required for all 15 nodes per integration cycle. This avoids redundant calculations and shortens the overall run time. It should be noted that, storing these matrix results consumes some memory, but this cost is often negligible compared to the computational time saved. This is particularly true when the computation of the CF and its gradient involves exponential functions, which are computationally expensive to evaluate repeatedly. 
\subsubsection{3. Internal information sharing} By analyzing the internal structure of the formulas, we also maximize the reuse of common intermediate results. A mathematical structure comparison of \eqref{eqBeta} and \eqref{eqGeneral} clearly reveals that the characteristic function and the exponential term constitute common components that appear in the numerical inversion formula. Therefore, in the algorithmic implementation, these shared computation results can be extracted and stored temporarily. This enables the reuse of these common results in subsequent calculations, rather than reevaluation. This refined information reuse mechanism avoids redundant computations, achieving efficient utilization of computational resources and improving the overall efficiency of the framework, while maintaining numerical accuracy. \par
\subsubsection{4. Contour deformation} To improve the integration of periodic functions, a contour deforming variable change is employed. The details are omitted for brevity.

In summary, through the aforementioned four optimization strategies, our method inherently ensures the computational efficiency required for solving optimization problems involving chance 
constraints.

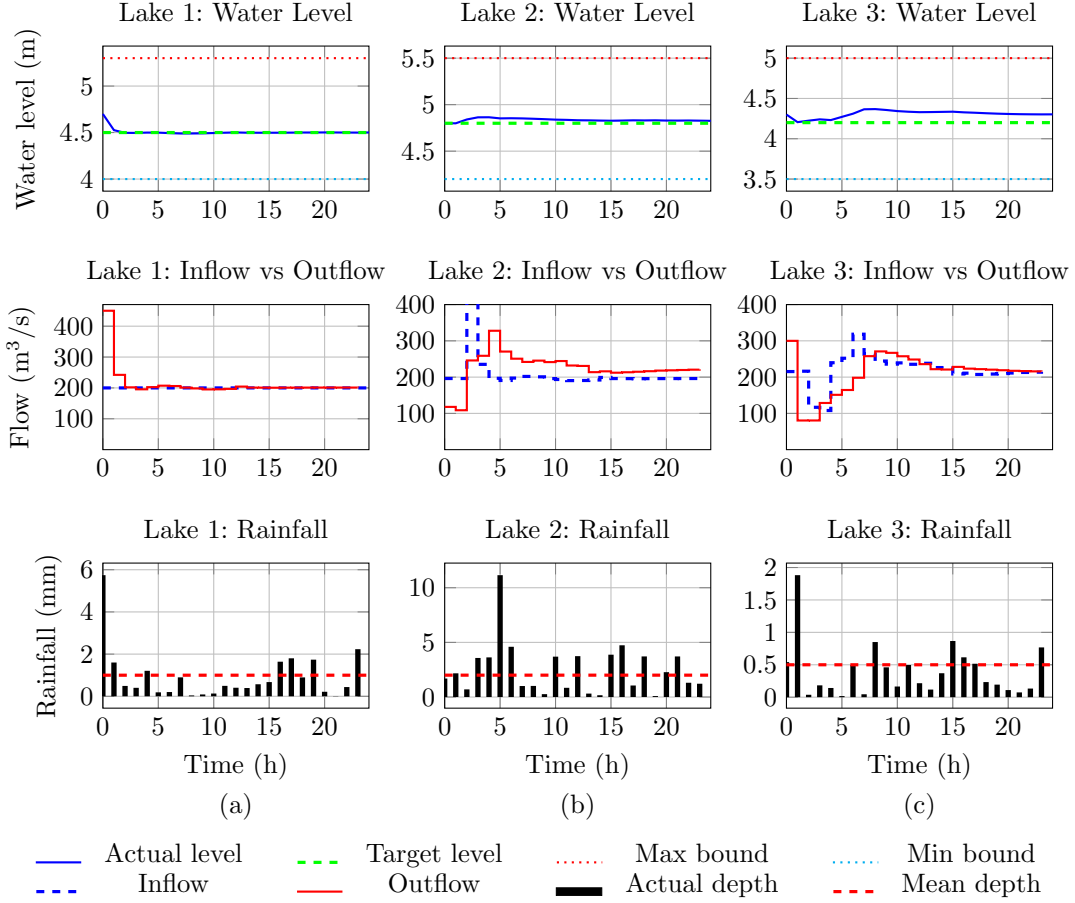
\begin{figure*}[!t]
\centering
\begin{tikzpicture}
\begin{groupplot}[
    group style={
        group size=3 by 3,
        horizontal sep=1.0cm,
        vertical sep=1.5cm,
    },
    width=0.28\textwidth,
    height=3.5cm,
    grid=both,
    xmin=0, xmax=24,
    xtick={0,5,10,15,20},
    every axis plot/.append style={mark=none},
]

\nextgroupplot[
    title={Lake 1: Water Level},
    ylabel={Water level (m)},
    xticklabels={0,5,10,15,20},
]
\addplot[lakeActual] table[x=tL,y=h1]{lakes.dat};
\addplot[lakeTarget] coordinates {(0,4.5) (24,4.5)};
\addplot[lakeMax]    coordinates {(0,5.3) (24,5.3)};
\addplot[lakeMin]    coordinates {(0,4.0) (24,4.0)};

\nextgroupplot[
    title={Lake 2: Water Level},
    xticklabels={0,5,10,15,20},
]
\addplot[lakeActual] table[x=tL,y=h2]{lakes.dat};
\addplot[lakeTarget] coordinates {(0,4.8) (24,4.8)};
\addplot[lakeMax]    coordinates {(0,5.5) (24,5.5)};
\addplot[lakeMin]    coordinates {(0,4.2) (24,4.2)};

\nextgroupplot[
    title={Lake 3: Water Level},
    xticklabels={0,5,10,15,20},
]
\addplot[lakeActual] table[x=tL,y=h3]{lakes.dat};
\addplot[lakeTarget] coordinates {(0,4.2) (24,4.2)};
\addplot[lakeMax]    coordinates {(0,5.0) (24,5.0)};
\addplot[lakeMin]    coordinates {(0,3.5) (24,3.5)};

\nextgroupplot[
    title={Lake 1: Inflow vs Outflow},
    ylabel={Flow (m$^3$/s)},
    xticklabels={0,5,10,15,20},
    ymin=0, ymax=470,             
    ytick={100,200,300,400},        
]
\addplot[lakeInflow] coordinates {(0,200) (24,200)};
\addplot[lakeOutflow] table[x=tU,y=u1]{lakes.dat};

\nextgroupplot[
    title={Lake 2: Inflow vs Outflow},
    xticklabels={0,5,10,15,20},
    ymin=0, ymax=400,             
    ytick={100,200,300,400},        
]
\addplot[lakeInflow] table[x=tU,y=q12]{lakes.dat};
\addplot[lakeOutflow] table[x=tU,y=u2]{lakes.dat};

\nextgroupplot[
    title={Lake 3: Inflow vs Outflow},
    xticklabels={0,5,10,15,20},
    ymin=0, ymax=400,             
    ytick={100,200,300,400},        
]
\addplot[lakeInflow] table[x=tU,y=q23]{lakes.dat};
\addplot[lakeOutflow] table[x=tU,y=u3]{lakes.dat};

\nextgroupplot[
    title={Lake 1: Rainfall},
    ylabel={Rainfall (mm)},
    xlabel={Time (h)},
]
\addplot[lakeDepth] table[x=tU,y=w1]{lakes.dat};
\addplot[lakeMean] coordinates {(0,1) (24,1)};

\nextgroupplot[
    title={Lake 2: Rainfall},
    xlabel={Time (h)},
]
\addplot[lakeDepth] table[x=tU,y=w2]{lakes.dat};
\addplot[lakeMean] coordinates {(0,2) (24,2)};

\nextgroupplot[
    title={Lake 3: Rainfall},
    xlabel={Time (h)},
]
\addplot[lakeDepth] table[x=tU,y=w3]{lakes.dat};
\addplot[lakeMean] coordinates {(0,0.5) (24,0.5)};

\end{groupplot}

\node[below=10mm] at (group c1r3.south) {(a)};
\node[below=10mm] at (group c2r3.south) {(b)};
\node[below=10mm] at (group c3r3.south) {(c)};

\end{tikzpicture}


\vspace{0.4em}
\begin{center}
\begin{tabular}{@{}cc@{\qquad}cc@{\qquad}cc@{\qquad}cc@{}}
  \makebox[0.5cm][l]{\tikz[baseline]{\draw[blue, thick] (0,0) -- (0.6,0);}}      & Actual level &
  \makebox[0.5cm][l]{\tikz[baseline]{\draw[green, dashed, very thick] (0,0) -- (0.6,0);}} & Target level &
  \makebox[0.5cm][l]{\tikz[baseline]{\draw[red, dotted, thick] (0,0) -- (0.6,0);}}       & Max bound   &
  \makebox[0.5cm][l]{\tikz[baseline]{\draw[cyan, dotted, thick] (0,0) -- (0.6,0);}}      & Min bound   \\
  \makebox[0.5cm][l]{\tikz[baseline]{\draw[blue, dashed, very thick] (0,0) -- (0.6,0);}} & Inflow      &
  \makebox[0.5cm][l]{\tikz[baseline]{\draw[red, thick] (0,0) -- (0.6,0);}}               & Outflow     &
  \makebox[0.5cm][l]{\tikz[baseline]{\draw[fill=black] (0,-0.05) rectangle (0.6,0.05);}} & Actual depth&
  \makebox[0.5cm][l]{\tikz[baseline]{\draw[red, dashed, very thick] (0,0) -- (0.6,0);}}  & Mean depth  \\
\end{tabular}
\end{center}

\caption{Water level changes, inflow changes, outflow changes and rainfall dynamics in three-lake reservoir system.}
\label{fig:three-lakes}
\end{figure*}

\section{Case study}
To validate the performance of our approach, we present an example of a three-lake reservoir water level control system. 

\subsection{Case setup and system dynamics}
The system consists of three interconnected reservoirs: Lake~1, Lake~2 and Lake~3. Lake~1 and Lake~2, as well as Lake~2 and Lake~3, are connected by underground rivers. The system dynamics at time step $k+1$ are modeled using a discrete-time mass-balance equation
\begin{equation}
    x_{k+1} = Ax_k + Bu_k + Gw_k + c_k,
\end{equation}
where $x_k=[h_{k,1},h_{k,2},h_{k,3},r_{k,12},r_{k,23}]^\top$ is the state, $h_{k,j}$ denotes the water level of Lake~$j$, $r_{k,ij}$ denotes the underground river flow between Lake~$i$ and Lake~$j$. $u_k=[u_{k,1},u_{k,2},u_{k,3}]^\top$ is the control input, $u_{k,j}$ denotes the controllable outflow from lake $j$. $w_k=[w_{k,1},w_{k,2},w_{k,3}]^\top$, $w_{k,j}$ denotes the stochastic rainfall for Lake~$j$, which is modeled as an exponentially distributed random variable. $c_k=[\frac{\Delta t}{S_1}(r_0-\varsigma_1),-\frac{\Delta t}{S_2}\varsigma_2,0,\varsigma_1,\varsigma_2]^\top$ is a constant term, $\Delta t$ is the sampling interval, $S_j$ is the surface area of Lake~$j$, $r_0$ is the inflow of Lake~1, $\varsigma_j$ denotes the base outflow of Lake~$j$ to the underground river. Matrix $G=[I_3;\mathbf{0}_{2\times3}]$, matrices $A$ and $B$ are shown as follows
\begin{equation}
A = \begin{bmatrix}
1 & 0 & 0 & 0 & 0 \\
0 & 1 & 0 & \frac{\Delta t}{S_2} & 0 \\
0 & 0 & 1 & 0 & \frac{\Delta t}{S_3} \\
0 & 0 & 0 & 0 & 0 \\
0 & 0 & 0 & 0 & 0
\end{bmatrix},
B = \begin{bmatrix}
-\frac{\Delta t}{S_1} & 0 & 0 \\
0 & -\frac{\Delta t}{S_2} & 0 \\
0 & 0 & -\frac{\Delta t}{S_3} \\
\eta_1 & 0 & 0 \\
0 & \eta_2 & 0 
\end{bmatrix},   
\end{equation}
where $\eta_j$ denotes the proportion of $u_{k,j}$ that reaches the underground river after accounting for seepage and evaporation losses. \par
Since we focus mainly on the water level control performance, the output equation can be formulated as 
\begin{equation}
    y_k=Cx_k,
\end{equation}
where $y_k=[h_{k,1},h_{k,2},h_{k,3}]^\top$ is the output, $C=[I_3,\mathbf{0}_{3\times2}]$. \par
Additionally, to reduce conservatism of the open loop solution, we employ an affine disturbance feedback strategy \citep{Lof:03}. The predicted input is parameterized as
\begin{equation}
    u_{k+l|k} = v_{k+l|k} + \sum_{i=0}^{l-1}L_{l,i}w_{k+i|k},
\end{equation}
where open loop control $v_{k+l|k}$ and disturbance feedback $L_{l,i}$ are decision variables, $N$ is the prediction horizon, and the summation only involves disturbances up to $k+l-1$, which preserves causality. For the choice of $L$, refer to \cite{Lof:03}. 

\subsection{Control objective and constraints}
The primary control objective is to mitigate flooding while maintaining operational water levels, in the presence of stochastic rainfall. Therefore, the controller aims to minimize a cost function that penalizes deviations from the reference water levels $y_{\text{ref}}$
\begin{equation}
    J =  \sum_{k=0}^{N-1} ||\mathbb{E}(y_{k}) - y_{\text{ref}}||^2 .
\end{equation} \par 
Because $h_{k,j}$ in the $y_{k}$ is affine in $w_{k,j}$, its expectation has a closed form.
Therefore, the objective is deterministic and directly optimized by YALMIP using the fmincon solver. The system is subject to several critical constraints:
\subsubsection{1. Flood safety constraint}There is a probability of $\gamma_1$ that water levels will not exceed the maximum flood capacities $y_{\max}$. This can be formulated as a chance constraint:
\begin{equation}
    P(y_{k,j} \le y_{\max,j}) \ge \gamma_1,\quad j=1,2,3. \label{consGamma1}
\end{equation}
\subsubsection{2. Drought safety constraint}The probability of water levels remaining above minimum levels $y_{\min}$ is $\gamma_2$, which leads to:
\begin{equation}
    P(y_{k,j} \geq y_{\min,j})\geq\gamma_2,\quad j=1,2,3. \label{consGamma2}
\end{equation}
\subsubsection{3. Water release constraint}The water releases $u_k$ is influenced by the rainfall $w_{k-1}$ and requires a case-specific constraint:
\begin{equation}
\begin{cases}
    0 \leq u_k \leq u_{\max}, \qquad\qquad w_{k-1}=0 \\
    P(0 \leq u_k \leq u_{\max}) \geq \gamma_3, \, \text{otherwise} \label{consGamma3}
\end{cases}
\end{equation}
It should be noted that the expression of $h_k$ also contains the stochastic rainfall $w_{k-1}$, thus the constraint in \eqref{consGamma1}, \eqref{consGamma2} and the second case in \eqref{consGamma3} can all be transformed to the standard form of \eqref{eqChanceCons}. This is a representative non-Gaussian case, which exposes the limitations of traditional SMPC methods: standard analytical reformulation yields inaccurate results due to distribution mismatch with the Gaussian distribution assumption; scenario-based approaches require massive samples to achieve reliable probability guarantees, thereby reducing computational feasibility. In contrast, our method only requires known distribution parameters to compute this chance constraint.

\subsection{Simulation and analysis}
According to the stochasticity in the constraint, the problem is formulated as a SMPC problem. The proposed computational framework is used to handle the chance constraint. The controller solves for an optimal sequence of control actions $u_{k}$ at each time step. The performance of this strategy is evaluated in a closed-loop simulation over $T=24\,\text{h}$. The remaining parameters are set to $N=10$, $\gamma_1=\gamma_2=\gamma_3=0.95$ and $\Delta t=1\,\text{h}$. \par
The results of the water level control system are shown in Fig.~\ref{fig:three-lakes}. \par
Fig.~\ref{fig:three-lakes} (a) shows the control performance for Lake~1. Water level control for Lake~1 demonstrates the most stable performance over the entire 24-hour horizon. The actual water level matches the target level almost exactly, and the rainfall remains consistently low and steady. With the inflow being fixed, the outflow closely tracks the inflow, indicating that the controller maintains an effective flow balance. The results demonstrate that under low and stable disturbances, the control system exhibits excellent stability. \par
Fig.~\ref{fig:three-lakes} (b) shows the control performance for Lake~2. Lake~2 experiences significant and persistent disturbances, yet the controller effectively suppresses water level fluctuations. The actual level remains slightly above the reference level, but is steadily approaching the reference value. Around $t=2$ h, there is a marked increase in inflow, resulting from the large release from Lake~1 at $t=0$. Meanwhile, rainfall over Lake~2 intensifies after $t=3$ h. The outflow responds promptly to both changes, increasing to match the higher inflow. Through rapid outflow adjustment, the controller counteracts the upward trend in the water level, maintaining stable operation within the safety bounds and ensuring constraint satisfaction. \par
Fig.~\ref{fig:three-lakes} (c) shows the control performance for Lake~3. Lake~3 exhibits the most irregular flow variations. The actual water level initially drops toward the reference level, then rises slightly, and eventually settles at a value marginally above the target level, while consistently remaining within the safety constraints. Rainfall is relatively heavy during the first two hours, leading to a gradual increase in outflow starting at $t=3$ h in order to adapt to both the changing inflow and the rainfall. The controller demonstrates strong adaptability, effectively balancing complex inflow fluctuations and rainfall effects through outflow adjustments, thereby maintaining the water level near the desired setpoint. \par

\section{Conclusion}
In stochastic model predictive control problems with non-Gaussian disturbances, this paper proposes a novel and unified framework that provides a widely applicable and efficient numerical computational scheme for chance constraints. Unlike methods relying on closed-form expressions of distributions or costly scenario sampling, our approach handles arbitrary single component distributions or mixture models within a unified structure. Specifically, this framework first utilizes the characteristic function of the disturbance distribution to compute the required probabilities and gradient via the Gil-Pelaez inversion formula. Then we employ the Gauss-Kronrod quadrature to reduce computational cost through vectorized integration points and reuse of precomputed information. A numerical example demonstrates the controller performs well under non-Gaussian disturbances.

\section*{DECLARATION OF GENERATIVE AI AND AI-ASSISTED TECHNOLOGIES IN THE WRITING PROCESS}
During the preparation of this work the authors used ChatGPT in order to assist with English sentence polishing. After using this tool, the authors reviewed and edited the content as needed and take full responsibility for the content of the publication.

\bibliography{ifacconf}             

@article{Qin:03,
  author  = {Qin, S. Joe and Badgwell, Thomas A.},
  title   = {A Survey of Industrial Model Predictive Control Technology},
  journal = {Control Engineering Practice},
  volume  = {11},
  number  = {7},
  pages   = {733--764},
  year    = {2003}
}

@phdthesis{Ras:14,
  author = {Halvgaard, Rasmus},
  title  = {Model Predictive Control for Smart Energy Systems},
  school = {Technical University of Denmark},
  year   = {2014}
}

@incollection{Bem:99,
  author    = {Bemporad, Alberto and Morari, Manfred},
  title     = {Robust Model Predictive Control: A Survey},
  booktitle = {Robustness in Identification and Control},
  editor    = {Garulli, Andrea and Tesi, Alberto and Vicino, Antonio},
  series    = {Lecture Notes in Control and Information Sciences},
  volume    = {245},
  pages     = {207--226},
  publisher = {Springer},
  address   = {London},
  year      = {1999}
}

@article{Joe:20,
  author  = {Paulson, Joel A. and Buehler, Edward A. and Braatz, Richard D. and Mesbah, Ali},
  title   = {Stochastic Model Predictive Control with Joint Chance Constraints},
  journal = {International Journal of Control},
  volume  = {93},
  number  = {1},
  pages   = {126--139},
  year    = {2020}
}

@inproceedings{Lof:03,
  author    = {L{\"o}fberg, Johan},
  title = {Approximations of closed-loop minimax {MPC}},
  booktitle = {Proceedings of the 42nd IEEE Conference on Decision and Control},
  volume    = {2},
  pages     = {1438--1442},
  year      = {2003}
}

@inproceedings{De:06,
  author    = {Mu{\~n}oz de la Pe{\~n}a, David and Alamo, Teodoro and Bemporad, Alberto and Camacho, Eduardo F.},
  title     = {Feedback Min-Max Model Predictive Control Based on a Quadratic Cost Function},
  booktitle = {Proceedings of the 2006 American Control Conference},
  address   = {Minneapolis, Minnesota, USA},
  year      = {2006},
  pages     = {1575--1580}
}

@inproceedings{Old:08,
  author    = {Oldewurtel, Frauke and Jones, Colin N. and Morari, Manfred},
  title     = {A Tractable Approximation of Chance Constrained Stochastic {MPC} Based on Affine Disturbance Feedback},
  booktitle = {Proceedings of the 47th IEEE Conference on Decision and Control},
  pages     = {4731--4736},
  year      = {2008}
}

@inproceedings{Bla:09,
  author    = {Blackmore, Lars and Ono, Masahiro},
  title     = {Convex Chance Constrained Predictive Control without Sampling},
  booktitle = {AIAA Guidance, Navigation, and Control Conference},
  address   = {Chicago, Illinois},
  year      = {2009},
  pages     = {7--21}
}

@article{Tan:24,
  author  = {Tan, Yuan and Yang, Jun and Chen, Wen-Hua and Li, Shihua},
  title   = {A Distributionally Robust Optimization Approach to Two-Sided Chance-Constrained Stochastic Model Predictive Control with Unknown Noise Distribution},
  journal = {IEEE Transactions on Automatic Control},
  volume  = {69},
  number  = {1},
  pages   = {574--581},
  year    = {2024}
}

@article{Geo:14,
  author  = {Schildbach, Georg and Fagiano, Lorenzo and Frei, Christoph and Morari, Manfred},
  title   = {The Scenario Approach for Stochastic Model Predictive Control with Bounds on Closed-Loop Constraint Violations},
  journal = {Automatica},
  volume  = {50},
  number  = {12},
  pages   = {3009--3018},
  year    = {2014}
}

@inproceedings{Kel:21,
  author    = {de Melo, Maisa Kely and Cardoso, Rodrigo T. N. and Jesus, Tales A.},
  title     = {A Genetic Algorithm for Investment Tracking with Stochastic Model Predictive Control},
  booktitle = {2021 IEEE Congress on Evolutionary Computation (CEC)},
  year      = {2021},
  pages     = {1543--1550}
}

@incollection{Bor:05,
  author    = {Borak, Szymon and H{\"a}rdle, Wolfgang and Weron, Rafa{\l}},
  title     = {Stable Distributions},
  booktitle = {Statistical Tools for Finance and Insurance},
  editor    = {{\v C}{\'i}{\v z}ek, Pavel and H{\"a}rdle, Wolfgang and Weron, Rafa{\l}},
  pages     = {21--44},
  publisher = {Springer},
  address   = {Berlin, Heidelberg},
  year      = {2005},
}

@book{Cas:02,
  author    = {Casella, George and Berger, Roger L.},
  title     = {Statistical Inference},
  edition   = {2},
  publisher = {Duxbury},
  address   = {Pacific Grove, CA},
  year      = {2002}
}

@article{Gil:51,
  author  = {Gil-Pelaez, J.},
  title   = {Note on the Inversion Theorem},
  journal = {Biometrika},
  volume  = {38},
  number  = {3-4},
  pages   = {481--482},
  year    = {1951}
}

@inproceedings{Lof:04,
  author    = {L{\"o}fberg, Johan},
  title     = {{YALMIP}: A Toolbox for Modeling and Optimization in {MATLAB}},
  booktitle = {Proceedings of the 2004 IEEE International Symposium on Computer Aided Control Systems Design},
  pages     = {284--289},
  year      = {2004}
}

@article{Lof:12,
  author    = {L{\"o}fberg, Johan},
  title     = {Automatic Robust Convex Programming},
  journal   = {Optimization Methods and Software},
  volume    = {27},
  number    = {1},
  pages     = {115--129},
  year      = {2012},
  publisher = {Taylor \& Francis}
}

@article{Bar:25,
  author  = {Barbosa, Filipe Marques and L{\"o}fberg, Johan},
  title   = {Exponential Cone Approach to Joint Chance Constraints in Stochastic Model Predictive Control},
  journal = {International Journal of Control},
  volume  = {98},
  number  = {12},
  pages   = {3024--3034},
  year    = {2025}
}








\end{document}